\documentclass[
longbibliography,
showpacs,
floatfix,
aps, 
prl,
twocolumn,
superscriptaddress,
amssymb,
]{revtex4-2}

\usepackage{amsmath}
\usepackage{tabulary}
\usepackage{tabularx}
\usepackage{url}
\usepackage[breaklinks=true]{hyperref}

\usepackage{graphicx}
\usepackage{times}
\usepackage{wasysym}
\usepackage{amssymb}
\usepackage{latexsym}

\usepackage{dcolumn}
\usepackage{amsfonts}
\usepackage{bm}
\usepackage{epsfig}

\newcommand{\be}{\begin{equation}}
\newcommand{\ee}{\end{equation}}
\newcommand{\bea}{\begin{eqnarray}}
\newcommand{\eea}{\end{eqnarray}}

\def\tcr{\tau_{\rm cr}}

\def\Edc{\mathcal{E}_j}
\def\Eac{\mathcal{E}}

\def\m{m^\star}
\def\eac{\epsilon}

\def\epseff{\varepsilon_{\mathrm{eff}}}

\def\oc{\omega_{\mbox{\scriptsize {c}}}}

\def\tq{\tau_{\rm q}}
\def\ttr{\tau}
\def\tem{\tau_{\rm em}}
\def\tsh{\tau_{\rm sh}}

\newcommand{\req}[1]{Eq.\,(\ref{#1})}

\newcommand{\rfig}[1]{Fig.\,\ref{#1}}

\newcommand{\rref}[1]{Ref.\,\onlinecite{#1}}
\newcommand{\rrefs}[2]{Refs.\,\onlinecite{#1},\,\onlinecite{#2}}

\def\ne{n_e}

\def\ne{n_e}

\begin{document}
\title{Radiowave-induced Resistance Oscillations}

\author{E. Bell}
\affiliation{School of Physics and Astronomy, University of Minnesota, Minneapolis, Minnesota 55455, USA}
\author{A. Zhakenuly}
\affiliation{School of Physics and Astronomy, University of Minnesota, Minneapolis, Minnesota 55455, USA}
\author{C. Hnatovsky}
\affiliation{National Research Council of Canada, Ottawa, Ontario K1A 0R6, Canada}
\author{K. W. Baldwin}
\author{L. N. Pfeiffer}
\author{K. W. West}
\affiliation{Department of Electrical Engineering, Princeton University, Princeton, New Jersey 08540, USA}
\author{S. Studenikin}
\affiliation{National Research Council of Canada, Ottawa, Ontario K1A 0R6, Canada}
\author{M. A. Zudov}
\email[Corresponding author: ]{zudov001@umn.edu}
\affiliation{School of Physics and Astronomy, University of Minnesota, Minneapolis, Minnesota 55455, USA}
\received{\today}

\begin{abstract}
Microwave-induced resistance oscillations (MIROs)  occur when a 2D electron gas is subjected to radiation of frequency $\omega = 2 \pi f$ and varying magnetic field $B$.
MIROs are periodic in $1/B$, with the period determined by the radiation frequency $\omega$, and their amplitude scales with the radiation power.
Stepping from single-photon transitions between Landau levels, MIROs are found on the low-field side of the cyclotron resonance, $\omega_c \lesssim  \omega$, where $\oc$ is the cyclotron frequency.
Here, we report on another class of magneto resistance oscillations, which are induced by high-intensity radiation in the radio frequency range and occur at $\omega_c \gg \omega$. 
These oscillations are independent of frequency $\omega$, can be either $1/B$ or $1/B^2$-periodic, and their period is controlled by the radiation electric field.
We further show that using a displacement model in the limit of short-range (``sharp'') disorder we can extract the radiation field and the width of the cyclotron resonance.
\end{abstract}

\maketitle
Discoveries of microwave-induced \cite{zudov:2001a}, phonon-induced \cite{zudov:2001b}, and Hall field-induced \cite{yang:2002} resistance oscillations opened a vibrant field of quantum transport in very high-Landau levels (LLs) of two-dimensional electron systems \cite{dmitriev:2012}. 
These initial findings were followed by observations of states with vanishing resistance \cite{mani:2002,zudov:2003}, conductance \cite{yang:2003}, and differential resistance \cite{bykov:2007,zhang:2008,hatke:2010a}. 
Microwave-induced resistance oscillations (MIROs) can be understood in terms of inter-LL electron transitions upon single-photon absorption \cite{durst:2003,vavilov:2004,dmitriev:2005}, i.e., in the regime of low radiation power.
As such, MIROs occur on the lower field side of the cyclotron resonance and the ratio of the radiation frequency $\omega = 2\pi f$ to the cyclotron frequency $\oc = eB/m^\star$, where $m^\star$ is the effective mass, is $\omega/\oc \gtrsim 1$.

In this Letter we focus on the regime of $\omega/\oc \ll 1$ which reveals another class of magneto-oscillations, which we term radiowave - induced resistance oscillations (RIROs). 
These oscillations are qualitatively distinct from MIROs in several aspects.
First, in contrast to MIROs, whose frequency is solely determined by $\omega$, the frequency of RIROs is independent of $\omega$, but is instead proportional to the radiation electric field $\Eac$.
Second, unlike MIRO amplitude, which scales with the radiation power, the amplitude of RIROs is power-independent.
Finally, RIROs can be periodic in either $1/B$ or in $1/B^2$, with the crossover between these two regimes occurring at $\oc \tcr = 1$, where $\tcr^{-1}$ is the width of the cyclotron resonance.
We show that the displacement contribution in the limit of sharp-range disorder \cite{vavilov:2007,khodas:2008,dmitriev:2009b,hatke:2011e,shi:2017a} captures main experimental features of RIROs.
We also highlight a striking similarity between RIROs and Hall field-induced resistance oscillations.

Our sample is a 60 $\mu$m-wide Hall bar, with a 200 $\mu$m separation between neighboring potential contacts, fabricated by photo-lithography \citep{hnatovsky:2022} from a 34.5 nm-wide GaAs/AlGaAs quantum well structure grown by molecular beam epitaxy. 
The sample was mounted in a $^3$He cryostat inside a superconducting solenoid with the magnetic field $B$ directed perpendicular to the sample. 
After low-temperature illumination, the electron density and mobility were $n_e \approx 3 \times10^{15}$ m$^{-2}$ and $\mu \approx 2 \times10^3$ m$^2$/Vs, respectively. 
Radiowaves were delivered to the sample via a CoBe semirigid coaxial cable terminated with a loop antenna mounted near the sample surface.
The longitudinal resistivity $\rho$ was measured using a standard four-terminal, low-frequency (a few Hz) lock-in technique.

\begin{figure}[t] 
\includegraphics{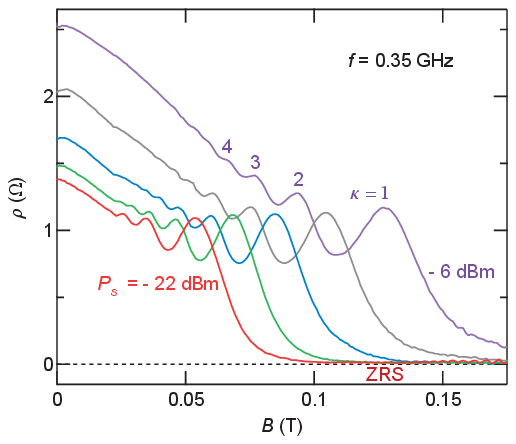}
\caption{\small{
Longitudinal resistivity $\rho$ versus magnetic field $B$ measured under radiowaves of frequency $f = 0.35$ GHz at different source powers from $P_s = -22$ dBm (left curve) to $P_s = -6$ dBm (right curve), in a step of 4 dBm.
The maxima of observed oscillations are marked by $\kappa = 1, 2, 3,\,...$ next to $\rho(B)$ measured at $P_s = -6$ dBm.
ZRS stands for zero-resistance state, developed at $B  > B_1$, where $B_1$ is the field of a fundamental RIRO maximum ($\kappa = 1$).
}}
\label{fig1}
\end{figure}

In \rfig{fig1} we present the longitudinal resistivity $\rho$ measured under radiowaves of frequency $f = 0.35$ GHz at different source powers from $P_s = -22$ dBm (left curve) to $P_s = -6$ dBm (right curve), in steps of 4 dBm, as a function of magnetic field $B$.
At all radiation powers, we observe magneto-resistance oscillations, which extend to higher $B$ with increasing $P_s$.
In what follows, we will refer to  these oscillations as RIROs.
At magnetic fields $B > B_1$, where $B_1$ is the field of the first RIRO maximum (marked by $\kappa = 1$ for $\rho(B)$ measured at $P_s = -6$ dBm in \rfig{fig1}), the resistivity decreases to nearly zero, signaling the formation of a zero-resistance state.
We also see the rise of both $\rho$ at $B = 0$ and of the minimum magnetic field at which oscillations start to develop with increasing $P_s$.
Both of these observations can be attributed to heating by the radiation; the higher zero-field resistivity and the higher onset of the oscillations are naturally explained by the increased electron-phonon \citep{stormer:1990} and electron-electron \citep{hatke:2009a,hatke:2009c} scattering, respectively \cite{note:heat}.

At the highest power, $P_s = - 6$ dBm, the first oscillation maximum, marked by $\kappa = 1$, occurs at $B_1\approx 0.13$ T.
At this field, electrons undergo transitions between two neighboring LLs separated by the cyclotron energy $\hbar \oc \approx 2.9$ K (0.25 meV).
Since the photon energy is only $\hbar\omega \approx 17$ mK (1.4 $\mu$eV), one concludes that electron transitions involve $N_{ph} \approx 1.7 \cdot 10^2$ photons.
The second maximum corresponds to a transition with the change in the LL index equal to 2.
It is found at $B_2 \approx 0.092$ T and involves $N_{ph} \approx 2.4 \cdot 10^2$ photons.
Since $B_1/B_2 \approx 1.4 < 2 $, we conclude that, in contrast to MIROs, these oscillations are not periodic in $1/B$. 
As we show below, at sufficiently high $B$, RIROs follow $1/B^2$ dependence (note that $B_1/B_2 \approx \sqrt{2}$).
At lower $B$, however, RIROs can become $1/B$-periodic, a regime which can be realized at lower $P_s$ \citep{bykov:2005c,mi:2019}.

\begin{figure}[t] 
\includegraphics{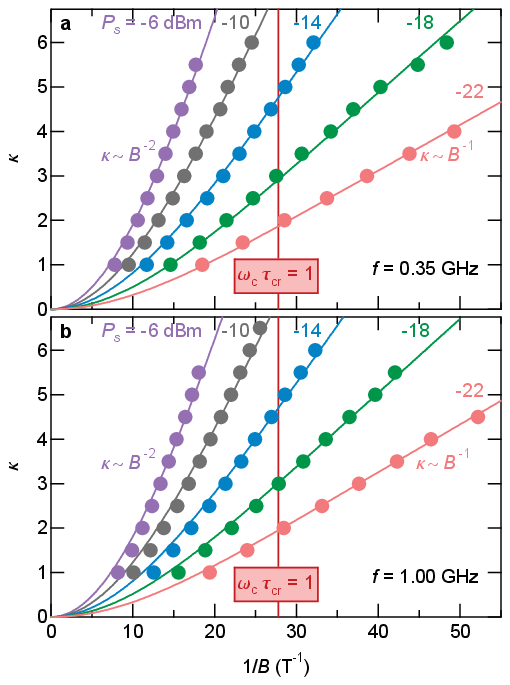}
\caption{\small{
Order of the oscillation maxima $\kappa$ (solid circles) as a function of inverse magnetic field $1/B$ obtained from the extrema of magnetoresistivity measured under radiowaves of frequency (a) $f = 0.35$ GHz and (b) $f = 1.00$ GHz at different source powers from $P_s = -6$ dBm (left) to $P_s = -22$ dBm (right), as marked.
Solid lines are fits to \req{eq.k}. 
Vertical line, drawn at $\oc\tcr = 1$, illustrates a crossover between the $\kappa \propto B^{-2}$, \req{eq.rf.h}, and $\kappa \propto B^{-1}$, \req{eq.rf.l}, regimes.
}}
\label{fig2}
\end{figure}
We next examine the RIRO periodicity in detail and, importantly, demonstrate that there is essentially no dependence on the radiowave frequency $\omega$.
To this end, in \rfig{fig2} we present a dimensionless parameter $\kappa$, which enumerates the oscillation maxima (integer) and minima  (half-integer), as a function of $1/B$ for (a) $f = 0.35$ GHz and (b) $f = 1.00$ GHz at different powers from $P_s = -6$ dBm (left) to $P_s = -22$ dBm (right), as marked.
It is obvious that the data presented in panels (a) and (b) are strikingly similar, despite different frequencies, strongly suggesting that frequency plays no significant role.
One also observes that the $\kappa$ dependencies are roughly parabolic at lower $1/B$, whereas at higher $1/B$ they become approximately linear.
As we explain below, the transition between these two regimes takes place when the cyclotron frequency becomes comparable to the width of the cyclotron resonance, as illustrated by the vertical line marked $\oc\tcr = 1$.
Here, $\tcr^{-1}$ characterizes the width of the cyclotron resonance, $\tcr^{-1} = \ttr^{-1} + \tem^{-1} \approx \tem^{-1}$, 
$\ttr$ is the momentum relaxation time,
$\tem^{-1} = \ne e^2/(2\sqrt{\epseff}\varepsilon_0\m c)$ is the radiative decay rate \citep{chiu:1976,zhang:2014}, $2\sqrt{\epseff}=\sqrt{\varepsilon}+1$, and $\varepsilon=12.8$ is the dielectric constant of GaAs.
The position of the $\oc\tcr = 1$ line in \rfig{fig2} is thus given by $B^{-1} e = \tcr/\m \approx \tem /\m e = (\sqrt{\varepsilon}+1)\varepsilon_0 c/\ne e$.

We next show that our experimental findings can be explained  by the  displacement contribution to photoresistance in the limit of sharp disorder \cite{vavilov:2007,khodas:2008,dmitriev:2009b,hatke:2011e,shi:2017a} which, under the conditions of our experiment ($\omega \ll \oc, \kappa \gtrsim 1$) can be expressed as \cite{zudov:2026,note:th}
\be
\frac{ \delta \rho_\omega }{ \rho_{D} } \approx \frac{8}\pi \frac {\tau} {\tsh} 
\lambda^2 \cos 2\pi \kappa\,,
\label{eq.rf}
\ee
where $\rho_{D}$ is the Drude resistivity, $\lambda = \exp(-\pi/\oc\tq)$ is the Dingle factor, $\tq$ is the quantum lifetime, $\tsh^{-1}$ is the sharp disorder scattering rate, 
\be
\kappa = \frac{e\Eac_\star (2R_c)}{\hbar \oc} \frac 1 {\sqrt{1+(\oc\tcr)^{-2}}}\,,
\label{eq.k}
\ee
$\Eac_\star = \Eac/\sqrt{\epseff}$, $\Eac$ is the radiation electric field, and $R_c$ is the cyclotron radius.
We see that $\kappa$ is \emph{independent} of $\omega$ and is determined by the electric field $\Eac$, magnetic field $B$, and $\tcr$.

It is also evident that $\tcr^{-1}$ plays an important role in the periodicity of oscillations.
At $\omega\tcr \gg 1$ \req{eq.k} simplifies to
\be
\kappa \approx  \frac{e \Eac_\star (2R_c)}{ \hbar \oc} \propto B^{-2}\,,
\label{eq.rf.h}
\ee
explaining the $1/B^2$ periodicity of the oscillations at higher $B$ seen in \rfig{fig2}.
On the other hand, when $\omega\tcr \ll 1$, \req{eq.k} reduces to
\be
\kappa \approx  \frac{e \Eac_\star (2R_c)}{ \hbar \tcr^{-1}} \propto B^{-1}\,,
\label{eq.rf.l}
\ee
leading to the conventional $1/B$ periodicity seen at lower $B$ in \rfig{fig2}.
In both regimes, the frequency of the oscillations scales with the electric field $\Eac$, accounting for the expansion of RIROs to larger $B$ with increasing $P_s$, as shown in \rfig{fig1}.

\begin{figure}[t] 
\includegraphics{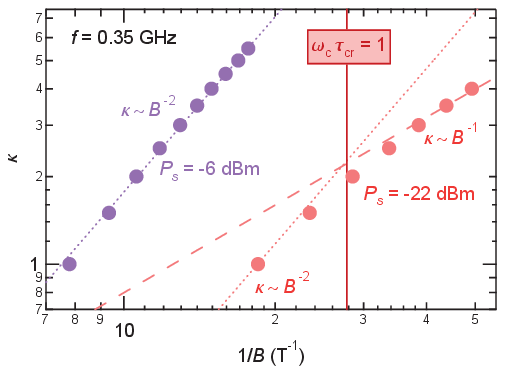}
\caption{\small{
$\kappa$ (solid circles) as a function of inverse magnetic field $1/B$ obtained from the extrema of magnetoresistivity measured under radiowaves of frequency $f = 0.35$ GHz at $P_s = -6$ dBm (left) and $P_s = -22$ dBm (right), as marked.
Dotted lines and dashed line represent fits to \req{eq.rf.h} and   \req{eq.rf.l}, respectively.  
Vertical line, drawn at $\oc\tcr = 1$, illustrates a crossover between the $\kappa \propto B^{-2}$ (\req{eq.rf.h}) and $\kappa \propto B^{-1}$ (\req{eq.rf.l}) regimes.
}}
\label{fig3}
\vspace{-0.15 in}
\end{figure}
\begin{figure}[b] 
\includegraphics{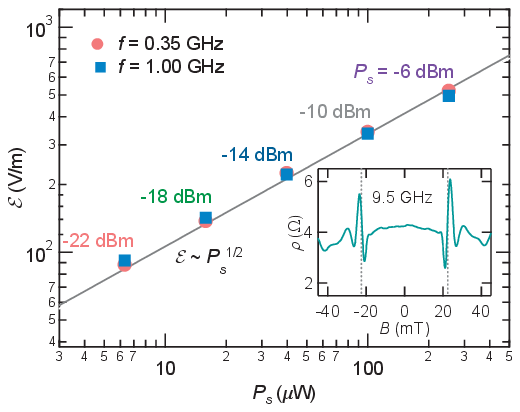}
\caption{\small{ 
Radiowave electric field $\Eac$, obtained from \req{eq.Eac}, as a function of the source power $P_s$ for $f = 0.35$ GHz (circles) and $f = 1.00$ GHz (squares).
Solid line illustrates the expected $\Eac \propto P_s^{1/2}$ dependence.
Inset shows $\rho$ versus $B$ measured at $f = 9.5$ GHz to obtain the effective mass $\m$ used to calculate $\Eac$.
}}
\label{fig4}
\vspace{-0.15 in}
\end{figure}
To further confirm the quasi-periodicity of the oscillations, we present in \rfig{fig3} $\kappa$ (solid circles) as a function of $1/B$ for $f = 0.35$ GHz at the highest power, $P_s = -6$ dBm, and at the lowest power, $P_s = -22$ dBm, as marked.
We indeed see that the high power data conform to $\kappa \propto 1/B^2$ dependence (dotted line) as they all fall in the region of $\oc\tcr > 1$, whereas the low power data clearly exhibit a crossover from $B^{-2}$ (dotted line) to $B^{-1}$ (dashed line) dependence occurring at $\oc\tcr = 1$ (vertical line).

We next examine the dependence of the radiation electric field $\Eac$, which we extract from our data, on the source power $P_s$.
For this purpose, we use the magnetic field at the primary maximum $B_1$, which remains at $\oc\tcr > 1$ over the whole range of $P_s$ studied.
Using \req{eq.rf.h}, we convert $B_1$ to
\be
\Eac = \sqrt{\frac{\epseff}{8\pi\ne}} \frac{e B_1^2}{\m}\,,
\label{eq.Eac}
\ee
and present the result in \rfig{fig4} as a function of $P_s$ for both $f = 0.35$ GHz and $f = 1.00$ GHz.
Here, we have used $\m = 0.06 m_0$, which was obtained from the higher frequency photoresistance data (see inset in \rfig{fig4}) and is in agreement with earlier measurements \cite{hatke:2013,fu:2017}.
As expected, we observe universal $\Eac \propto \sqrt{P_s}$ dependence over the whole range of $P_s$ for both frequencies.

\begin{figure}[t] 
\includegraphics{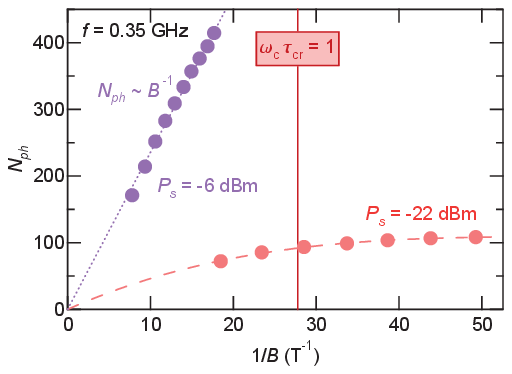}
\caption{\small{
Number of photons $N_{ph} = \kappa\oc/\omega$ (solid circles) as a function of inverse magnetic field $1/B$ obtained from the extrema of magnetoresistivity measured under radiowaves of frequency $f = 0.35$ GHz at $P_s = -6$ dBm and $P_s = -22$ dBm, as marked.
Dotted line illustrates $N_{ph} \propto B^{-1}$ dependence expected for high power data at $\oc\tcr > 1$, whereas the lower power data cross over to the expected saturation at $\oc\tcr < 1$ (the dashed line is a guide to an eye), see \req{eq.nph}. 
Vertical line is drawn at $\oc\tcr = 1$.
The cyclotron resonance takes place at $1/B \approx 1.3 \cdot 10^3$ T$^{-1}$.
}}
\label{fig5}
\vspace{-0.15 in}
\end{figure}
It is worthwhile to examine the number of participating photons, $N_{ph} = \kappa \oc/\omega$ which, in the limiting cases, can be approximated as
\be
N_{ph} \approx 
\frac{e \Eac^\star (2R_c)}{\hbar \omega} \cdot
\begin{cases} 
      1 \propto B^{-1}, & \oc\tcr \gg 1 \\
      \oc\tcr \propto B^{0}, & \oc\tcr \ll 1
   \end{cases}
\,.
\label{eq.nph}
\ee
In \rfig{fig5} we show $N_{ph}$ (solid circles) as a function of $1/B$ corresponding to the extrema of magnetoresistivity measured under radiowaves of frequency $f = 0.35$ GHz at $P_s = -6$ dBm and $P_s = -22$ dBm, as marked.
In agreement with \req{eq.nph}, the dotted line illustrates $N_{ph} \propto B^{-1}$ dependence expected for the high power data at $\oc\tcr \gg 1$.
The low power dependence, however, crosses over to a saturation regime, anticipated at $\oc\tcr \ll 1$ (the dashed line is a guide to an eye). 
The saturation value $N_{ph}^s$ is determined by three parameters, $n_e$, $\Eac$, and $\omega$, namely $N_{ph}^s = 4\sqrt{\varepsilon_0/\mu_0}\sqrt{2\pi/n_e}\Eac/e\omega$.
This expression offers another way to obtain the radiation electric field; with $N_{ph}^s \approx 110$, we obtain $\Eac \approx 80$ V/m, in good agreement with the lowest power data point in \rfig{fig4}.
While the importance of multi-photon processes has been noticed in multiple papers \cite{zudov:2006a,lei:2006b,khodas:2010,hatke:2011e,lei:2011, shi:2017a}, experimental evidence remained limited to processes with a number of participating photons $N_{ph} < 10$ and the regime of $\eac \sim 1$.

Finally, as mentioned in \rref{zudov:2026} there is a striking similarity between RIROs and Hall field-induced resistance oscillations \citep{yang:2002,zhang:2007a,zhang:2007b} which appear in the differential resistivity $r$ \citep{vavilov:2007},
\be
\frac{ \delta r_j }{ \rho_{D} } \approx \frac{16}{\pi} \frac {\tau}{\tsh} \lambda^2 \cos (2\pi \epsilon_j)\,,~\epsilon_j \approx  \frac{e \Edc (2R_c)}{\hbar \oc}\,,
\label{eq.hiro}
\ee
when a direct current $j$ is passed through a Hall bar-shaped sample creating Hall electric field $\Edc \propto B$.
We note that this phenomenon has also been observed in Corbino rings in which $\Edc$ is $B$-independent \cite{bykov:2012}.
This similarity aligns with the fact that the radiowave electric field $\Eac$ appears essentially static for electrons when $\omega \ll \oc$. 
In addition, the zero-resistance state at $B > B_1$ and its evolution with radiation power, \rfig{fig1}, are very similar to those associated with zero-differential resistance states observed under pure dc excitation \citep{zhang:2008,hatke:2010a, note:chep}.
It is interesting to investigate RIROs under dc electric fields, as was done for MIROs \citep{zhang:2007c,hatke:2008a,hatke:2008b}, which would allow to further reinforce the interpretation of our findings. 

In summary, we have observed a new class of radiation-induced magnetoresistance oscillations which, in contrast to MIROs, emerge on the high-field side of the cyclotron resonance ($\oc \gg \omega$) under high intensity radiowaves.
We were able to explain our findings in terms of the displacement mechanism considering sharp disorder contribution.
The unique properties of these oscillations allow access to the radiowave electric field and the radiative decay rate, which governs the width of the cyclotron resonance.
Owing to the generic nature of the phenomenon, RIROs should be observable in a variety of 2D systems \cite{konstantinov:2009,zudov:2014,shchepetilnikov:2016,karcher:2016,otteneder:2018,monch:2020,savchenko:2020}. 
In addition, it should be relevant to Floquet physics, see e.g., \rrefs{khodas:2010}{shi:2024}.

\begin{acknowledgments}
M.A.Z. thanks I. Dmitriev, M. Dyakonov, D. Polyakov, Q. Shi, B. Shklovskii, I. Sodemann, and, especially, I. Gornyi for discussions.
The Princeton portion of this work is funded in part by the Gordon and Betty Moore Foundation EPiQS initiative, Grant GBMF14260. 
\end{acknowledgments}


\end{document}